\documentstyle[pra,multicol,aps,epsfig]{revtex}
\begin{document}
\renewcommand{\textfraction}{0.10} 
\renewcommand{\topfraction}{1.0} 
\renewcommand{\bottomfraction}{1.0} 
\flushbottom
\title{Dynamics of Atom-Mediated Photon-Photon Scattering I: Theory}
\author{M. W. Mitchell and R. Y. Chiao}
\address{Department of Physics, University of California at Berkeley,
Berkeley, CA 94720, USA  }
\date{\today} 
\maketitle

\begin{abstract}
The mediated photon-photon interaction due to the resonant Kerr 
nonlinearity in an inhomogeneously broadened atomic vapor is 
considered.  The time-scale for photon-photon scattering is computed 
and found to be determined by the inhomogeneous broadening and the 
magnitude of the momentum transfer.  This time can be shorter than the 
atomic relaxation time.  Effects of atom statistics are included and 
the special case of small-angle scattering is considered.  In the 
latter case the time-scale of the nonlinear response remains fast, 
even though the linear response slows as the inverse of the momentum 
transfer.  
\end{abstract}


\begin{multicols}{2}
\section{Introduction}

Recently there has been experimental and theoretical interest in the 
nonlinear optics of confined light \cite{Boyce=1999a}.  A medium 
possessing an optical Kerr nonlinearity and confined within a planar or 
cylindrical Fabry-Perot resonator gives rise to new nonlinear optical 
phenomena such as soliton filtering and bilateral symmetry breaking 
\cite{Boyce=1999b,Torres=1999}.  The classical nonlinear optics of 
this system is described by the Complex Ginzburg-Landau equation 
(CGLE)
\begin{eqnarray}
\label{CGLE}
\frac{\partial E}{\partial t} &=& \frac{i c}{2 n_{0} k} 
\nabla^{2}_{\perp}E
+ i\omega A \frac{n_{2}}{n_{0}}|E|^{2}E + \frac{ic\Delta k}{n_{0}}E
\nonumber \\
& & 
-\Gamma(E-E_{\rm d}),
\end{eqnarray} 
where $E$ is the electric field envelope, $k$ is the longitudinal 
wavenumber, $\omega=ck/n_{0}$ is the field envelope angular frequency, 
$A$ is a mode overlap factor, $\Delta k$ is the wavenumber mismatch 
from the linear-cavity response and $\Gamma$ is the field amplitude 
decay rate.  The classical dynamics of Eq. (\ref{CGLE}) describes 
the mean-field behavior of a system of interacting photons coherently 
coupled to an external reservoir.  A photonic system of this sort is a 
versatile model system for condensed matter physics in reduced 
dimensions \cite{AkhmanovMemorial}, as the parameters $\Delta 
k$,$n_{2}$, $\Gamma$, and $E_{\rm d}$ in Eq. (\ref{CGLE}) are subject 
to experimental control.  In particular, an atomic vapor can provide a 
strong Kerr nonlinearity which is tunable both in strength and in 
sign.  In this case the nonlinearity arises from the saturation of the
linear refractive index, which is a strong function of the drive laser 
frequency near an absorption resonance.


Some of the most interesting proposed experiments for this system, 
including generation of few-photon bound states \cite{Deutsch=1992}, 
direct observation of the the Kosterlitz-Thouless transition in an 
optical system \cite{AkhmanovMemorial} and observation of quantum 
corrections to the elementary excitation spectrum of a 1D photon gas 
\cite{Lieb=1963,Lieb=1963a} intrinsically involve photon correlations.  
For this reason, it is important to understand the microscopic (and 
not just mean-field) behavior of photons in an optical Kerr medium.  
We specifically consider saturation of the resonant electronic 
polarization of a Doppler-broadened atomic vapor, a medium which has 
been proposed for quantum cavity nonlinear optics experiments and used 
to observe a nonlinear cavity mode \cite{Boyce=1999b}.  Thus the 
system under consideration involves dispersion, loss, inhomogeneous 
broadening, and the continuum of transverse modes in an extended 
resonator.

Sophisticated techniques have been developed for treating mediated 
interactions among photons in nonlinear media.  One approach is to 
obtain an effective theory in which the quanta are excitations of 
coupled radiation-matter modes, by canonical quantization of the 
macroscopic field equations \cite{Drummond=1987,Carter=1987}, or by 
direct attack on a microscopic Hamiltonian \cite{Drummond=1999}.  This 
approach has the advantage of generality and is suited to multi-mode 
problems, but has basic difficulties with loss and dispersion near 
resonance \cite{Hillery=1984,Drummond=1990,Abram=1991}.  Microscopic 
treatments include Scully-Lamb type theory 
\cite{Scully=1967,Sargent=1985} and application of phase-space methods 
\cite{Drummond=1981,Louisell=1973}.  A strength of these techniques is 
their ability to handle relaxation and population changes.  They are, 
however, cumbersome to apply to inhomogeneously broadened media and to 
multi-mode problems.

In this paper we characterize the atom-mediated photon-photon 
interaction using an accurate microscopic model and perturbation 
calculations.  This allows us to determine the time-scale of the 
mediated photon-photon interaction in the atomic vapor, despite the 
complexity of the medium.  We find that the interaction is fast and 
not intrinsically lossy, even for small momentum transfer.  Thus the 
medium is suitable for quantum optical experiments, including 
experiments using the NLFP as a model for the interacting Bose gas.

\section{Scattering Calculations}

The complete system is treated as the quantized electromagnetic field 
interacting via the dipole interaction with an vapor of atoms of mass 
$M$.  The perturbation calculations are performed in momentum space, as is 
natural for thermodynamic description of the atomic vapor.  This also 
makes simple the inclusion of atomic recoil effects.  The dipole 
interaction term is identified as the perturbation, so that the 
eigenstates of the unperturbed Hamiltonian are direct products of 
Fock states for each field.  In the rotating wave approximation, the 
unperturbed and perturbation Hamiltonians are
\begin{equation}
H_{0} = \sum_{{\bf k},\alpha}\hbar c k 
a_{{\bf k},\alpha}^{\dag}a_{{\bf k},\alpha}
+ \sum_{n,{\bf p}}(\hbar \omega_{n} + \frac{\hbar^{2}{\bf p}^{2}}{2 M})
c_{n,{\bf p}}^{\dag}c_{n,{\bf p}}
\end{equation}
\begin{eqnarray}
H' &= &  -{\bf E}({\bf x})\cdot {\bf d}(\bf{x})
\nonumber \\
& =&
-\sum_{{\bf k},\alpha}\sqrt{\frac{2 \pi \hbar c k}{V}} 
\sum_{n,m,{\bf p}} i{\bf e}_{{\bf k},\alpha}\cdot {\mbox{\boldmath $\mu$}}_{nm}
c_{n,{\bf p}+{\bf k}}^{\dag}c_{m,{\bf p}}a_{{\bf k},\alpha} 
\nonumber \\
& & + {\rm h.c.}
\end{eqnarray}
where $a_{{\bf k},\alpha}$ is the annihilation operator for a photon of 
momentum $\hbar{\bf k}$ and polarization $\alpha$, $c_{n,{\bf p}}$ is the 
annihilation operator for an atom in internal state $n$ with center-of-mass 
momentum $\hbar {\bf p}$, ${\bf E}$ is the quantized electric field and 
${\bf d}$ is the atomic dipole field.  Polarization plays only a 
very minor role in this discussion so polarization indices will be 
omitted from this point forward.


The simplest mediated interaction is photon-photon scattering, which 
transfers momentum from one photon to another by temporarily 
depositing this momentum in the medium.  Specifically, photons with 
momenta ${\bf k},{\bf l}$ are consumed and photons with momenta ${\bf 
k}'\equiv {\bf k}+{\bf q},{\bf l}'\equiv {\bf l}-{\bf q}$ are 
produced.  The lowest-order processes to do this are fourth order, so 
we look for relevant terms in $H'H'H'H'$.  A parametric process, i.e., 
one which leaves the medium unchanged, sums coherently over all atoms 
which could be involved \cite{Heidmann=1987}.  Due to this coherence, the rates of 
parametric processes scale as the square of $N/V$, the number density 
of atoms.  In contrast, incoherent loss processes such as Rayleigh and 
Raman scattering scale as $N/V$.  Thus for large atomic densities, a 
given photon is more likely to interact with another photon than it is 
to be lost from the system.  

In this sense, the interaction is not intrinsically lossy, as are some 
optical Kerr nonlinearities such as optical pumping or thermal 
blooming.  The latter processes require absorption of photons before 
there is any effect on other photons.  For this reason, they are 
unsuitable for quantum optical experiments such as creation of a 
two-photon bound state.

One parametric process, photon-photon scattering at a single 
atom, is described by the diagram of Fig. 1.
The 
relevant terms in $H'H'H'H'$ contain
\begin{eqnarray}
\label{OneAtomProcess}
&&c_{a,{\bf p}}^{\dag}c_{d,{\bf p}+{\bf l}'}a_{{\bf l}'}^{\dag} 
  c_{d,{\bf p}+{\bf l}'}^{\dag}c_{c,{\bf p}-{\bf q}}a_{{\bf l}}
\nonumber \\
&&\times c_{c,{\bf p}-{\bf q}}^{\dag}c_{b,{\bf p}+{\bf k}}a_{{\bf k}'}^{\dag}
  c_{b,{\bf p}+{\bf k}}^{\dag}c_{a,{\bf p}}a_{{\bf k}}
\end{eqnarray}
or permutations ${\bf k}' \leftrightarrow {\bf l}'$, 
${\bf k} \leftrightarrow {\bf l}$ for a total of four terms.
Here ${\bf p}$ is the initial atomic momentum and
$a$ through $d$ index the atomic states involved.
With the assumption that no atoms 
are initially found in the upper states $b$ and $d$, i.e., 
$n_{b}=n_{d}=0$, this reduces to
\begin{equation}
\label{OneAtom}
n_{a,{\bf p}}(1\pm n_{c,{\bf p}-{\bf q}})
a_{{\bf l}'}^{\dag}a_{{\bf l}}a_{{\bf k}'}^{\dag}a_{{\bf k}}
\end{equation}
where the $n$ are 
number operators for the atomic modes and the 
upper and lower signs hold for Bose and Fermi gases, 
respectively.  The difference for atoms of different statistics 
reflects the fact that the scattering process takes the atom through 
an intermediate momentum state which could be occupied.  Occupation 
of this intermediate state enhances the process 
for Bose gases but suppresses it for Fermi gases.

A thermal average of the relevant terms in $H'H'H'H'$ gives the 
thermally averaged effective perturbation
\begin{equation}
\left< H'_{\rm eff} \right>
= \frac{(2\pi)^{3}}{V}
\sum_{{\bf k}{\bf l}{\bf k}'{\bf l}'} 
V_{{\bf l}'{\bf k}'{\bf l}{\bf k}}
a_{{\bf l}'}^{\dag}a_{{\bf l}}a_{{\bf k}'}^{\dag}a_{{\bf k}}
\end{equation}
where
\begin{equation}
    \label{OneAtomEff}
V_{{\bf l}'{\bf k}'{\bf l}{\bf k}}
 \equiv 
\sum_{a} \int d^{3}{\bf p} v_{\rm eff}({\bf p},a,c)
\left<n_{a,{\bf p}}\right> 
\sum_{c} \left(1\pm \left<n_{c,{\bf p}-{\bf q}}\right>\right),
\end{equation}
\begin{equation}
v_{\rm eff}({\bf p},a,c) = v_{\rm eff}^{(1)}+ v_{\rm eff}^{(2)}+ 
v_{\rm eff}^{(3)}+ v_{\rm eff}^{(4)}
\end{equation} 
\begin{eqnarray}
\label{Geometrics}
v_{\rm eff}^{(1)} &=& \frac{c^{2}\sqrt{k l k' l'}}{(2 \pi)^{4} \hbar}
\nonumber \\
& & \times \sum_{bd} 
({\bf e}_{{\bf l}'}\cdot{\mbox{\boldmath $\mu$}}_{da})^{*}
{\bf e}_{{\bf l}}\cdot{\mbox{\boldmath $\mu$}}_{dc}
({\bf e}_{{\bf k}'}\cdot{\mbox{\boldmath $\mu$}}_{bc})^{*}
{\bf e}_{{\bf k}}\cdot{\mbox{\boldmath $\mu$}}_{ba}
\nonumber \\
& & \times
\left[R_{1}^{(1)} R_{2}^{(1)} R_{3}^{(1)}\right]^{-1}
\end{eqnarray}
and similar expressions obtain for $v_{\rm eff}^{(2-4)}$.  and 
$\left<n_{a,{\bf p}}\right>$ is the average occupancy of the atomic 
state $\left|a,{\bf p}\right>$.  The $R_{i}^{(1)}$ are the resonance 
denominators
\begin{eqnarray}
\label{ResDenomsOne}
R^{(1)}_{1} & = &
{c(k+l-k') -\frac{\hbar}{M}[{\bf p} \cdot {\bf l}' + l'^{2}/2]
- \omega_{da} + i \gamma_{d}}
\nonumber \\
R^{(1)}_{2} & = &
{c(k-k') - \frac{\hbar}{M}[-{\bf p} \cdot {\bf q} + q^{2}/2]
- \omega_{ca} + i \eta}
\nonumber \\
R^{(1)}_{3} & = &
{c(k) -\frac{\hbar}{M}[{\bf p} \cdot {\bf k} + k^{2}/2]
- \omega_{ba} + i \gamma_{b}}.
\end{eqnarray}
Here $\hbar \omega_{ij} \equiv \hbar(\omega_{i} - \omega_{j})$ is the 
energy difference between states $i$ and $j$, $\gamma_{i}$ is the 
inverse lifetime of state $i$ and $\eta$ is a vanishing positive 
quantity.  Here and throughout, the process is understood to conserve 
photon momentum, but for clarity of presentation this is not 
explicitly indicated.

As described in Appendix \ref{Correlations}, intensity correlation 
functions for photon-photon scattering products contain a
Fourier transform of the scattering amplitudes
\begin{equation}
P(x_{\rm A},t_{\rm A},x_{\rm B},t_{\rm B})
\propto
\left|\int d \delta_{k'} V_{{\bf l}'{\bf k}'{\bf l}_{0}{\bf k}_{0}} 
\exp[i c \delta k'\tau_{-}]\right|^{2}
\end{equation} 
where $\delta{k'}$ is the output photon energy shift, $x_{\rm A,B}$ 
and $t_{\rm A,B}$ are detection positions and times, respectively, and 
$\tau_{-} \equiv t_{\rm B}-x_{\rm B}/c-t_{\rm A}+x_{\rm A}/c$
is the difference in 
retarded times.  This expression 
allows us to determine the time correlations for photon-photon 
scattering in a number of important cases.

\section{Large-angle scattering}

The simplest configuration to understand is that of counterpropagating 
input beams producing counterpropagating output photons scattered at 
large angles.  This is also the most convenient experimental geometry.

\subsection{One Atom Process}

%
%

Scattering amplitudes and rates for right-angle scattering by the 
one-atom process are shown in Fig.  2
and Fig.  3, respectively.  For the moment we ignore the 
statistical correction due to the $n_{a,{\bf p}}n_{c,{\bf p}-{\bf q}}$ 
term in Eq. (\ref{OneAtomEff}), which will be considered separately.  
The the vapor is treated as a gas of two-level atoms.  The parameters 
are the Doppler width $\delta_{\rm D} \equiv k (k_{\rm B} T/M)^{1/2}$, 
where $k_{\rm B}$ is Boltzmann's constant, 
the radiative linewidth $\gamma_{b} = A_{\rm E}/2$ where $A_{\rm E}$ 
is the Einstein A coefficient, and the detuning $\Delta \equiv 
ck-\omega_{ba}$, in the ratios $\gamma_{b} = 0.01 \delta_{\rm D}$, 
$\Delta = 2 \pi \delta_{\rm D}$.  The amplitude units are arbitrary, 
but do not vary between graphs.

At this point it is important to note that the duration of the 
correlation signal is much shorter than the coherence lifetime of an 
individual atom, approximately $\gamma_{b}^{-1}$.  In fact, the 
duration of the correlation signal is determined by the momentum 
distribution, a property of the medium as a whole.  This can be 
explained in terms of the coherent summation of amplitudes for 
scattering processes occurring at different atoms.  The process is 
coherent only when it is not possible, even in principle, to tell 
which atom participated.  This clearly requires momentum conservation 
among the photons, but it also limits the duration of the atomic 
involvement.  An atom acting as intermediary to transfer momentum 
${\bf q}$ is displaced during the time in remains in the state $c$ of 
Fig. 1.
If this displacement is larger than the 
thermal deBroglie wavelength $\Lambda$ it is possible, in principle, to 
determine which atom participated.  This limits the duration of the 
coherent process to $\delta \tau \sim \Lambda M/\hbar q$.

\subsection{Statistical Correction}

%
As noted above, the quantum statistics of the atoms in the vapor 
contribute a correction to the single-atom scattering amplitude.  
This correction (with the sign appropriate for Bose atoms) is shown in 
Fig. 4
for a gas with phase space density
$N\Lambda^{3}/V = 1/2$, where $\Lambda \equiv (M k_{\rm 
B}T/2\pi\hbar^{2})^{1/2}$ is the thermal deBroglie wavelength.  
Parameters are as for Fig. 2.

\subsection{Simultaneous scattering}

%
A second parametric process, simultaneous scattering, 
is described by the diagram of Fig. 5.
The 
relevant terms in $H'H'H'H'$ contain
\begin{eqnarray}
&&c_{a,{\bf p}}^{\dag}c_{d,{\bf p}+{\bf l}'}a_{{\bf l}'}^{\dag} 
  c_{c,{\bf p}-{\bf q}}^{\dag}c_{b,{\bf p}+{\bf k}}a_{{\bf k}'}^{\dag}
\nonumber \\
&&\times c_{d,{\bf p}+{\bf l}'}^{\dag}c_{c,{\bf p}-{\bf q}}a_{{\bf l}}
  c_{b,{\bf p}+{\bf k}}^{\dag}c_{a,{\bf p}}a_{{\bf k}}
\end{eqnarray}
or permutations ${\bf k}' \leftrightarrow {\bf l}'$, 
${\bf k} \leftrightarrow {\bf l}$ for a total of four terms.
Making the same assumption as before, this reduces to
\begin{equation}
\label{TwoAtom}
n_{a,{\bf p}}n_{c,{\bf p}-{\bf q}} 
a_{{\bf l}'}^{\dag}a_{{\bf k}'}^{\dag}a_{{\bf l}}a_{{\bf k}}.  
\end{equation} 
This process corresponds to the absorption of each photon by an atom 
before emission of either, and thus describes a two-atom process and 
is of the same order in the atomic number density as the Fermi and 
Bose corrections to single-atom scattering.
The kinematical and geometric factors of Eq. (\ref{OneAtomEff}) 
and Eq. (\ref{Geometrics}) are the same for this process, and the 
resonance denominators are
\begin{eqnarray}
\label{ResDenomsTwo}
R^{(2)}_{1} & = &
c(k+l-k') -\frac{\hbar}{M}[{\bf p} \cdot {\bf l}' + l'^{2}/2]
- \omega_{da} + i \gamma_{d}
\nonumber \\
R^{(2)}_{2} & = &
c(k+l) - \frac{\hbar}{M}[{\bf p} \cdot {\bf k} + k^{2}/2
+({\bf p}-{\bf q}) \cdot {\bf l} + l^{2}/2]
\nonumber \\
& & - \omega_{ba} + i \gamma_{b} - \omega_{dc} + i \gamma_{d} 
\nonumber \\
R^{(2)}_{3} & = &
c(k) -\frac{\hbar}{M}[{\bf p} \cdot {\bf k} + k^{2}/2]
- \omega_{ba} + i \gamma_{b}.
\end{eqnarray}


Amplitudes for simultaneous scattering are shown in Fig. 6
for a gas with a phase space density of one half.
Parameters are as for Fig. 2.

\subsection{Fermi and Bose Gases}

%

The statistical correction and two-atom scattering contributions add 
coherently, giving considerably different correlation functions for 
moderate degeneracy Bose vs. Fermi gases.  This is illustrated in 
Fig. 7
and Fig. 8,
which show the scattering 
rates vs. delay for Bose and Fermi gases with a phase space density of 
one half.  Parameters are as for Fig. 2.

\subsection{Ladder Process}


In atoms with a ``ladder'' level 
structure, in which three levels $a$--$c$ are ordered in energy
$\omega_{c}>\omega_{b}>\omega_{a}$ and connected by matrix elements
$\mu_{ba},\mu_{cb}\neq 0$, $\mu_{ca}=0$, an additional 
process described by the diagram of Fig. 9
is 
possible.  The relevant terms in $H'H'H'H'$ contain
\begin{eqnarray}
\label{LadderProcess}
&&c_{a,{\bf p}}^{\dag}c_{d,{\bf p}+{\bf l}'}a_{{\bf l}'}^{\dag} 
  c_{d,{\bf p}+{\bf l}'}^{\dag}c_{c,{\bf p}+{\bf k}+{\bf l}}a_{{\bf k}'}^{\dag}
\nonumber \\
&&\times c_{c,{\bf p}+{\bf k}+{\bf l}}^{\dag}c_{b,{\bf p}+{\bf k}}a_{{\bf l}}
  c_{b,{\bf p}+{\bf k}}^{\dag}c_{a,{\bf p}}a_{{\bf k}}
\end{eqnarray}
or permutations ${\bf k}' \leftrightarrow {\bf l}'$, 
${\bf k} \leftrightarrow {\bf l}$ for a total of four terms.
Making the same assumption as before, this reduces to
\begin{equation}
\label{LadderAtom}
n_{a,{\bf p}}
a_{{\bf l}'}^{\dag}a_{{\bf k}'}^{\dag}a_{{\bf l}}a_{{\bf k}}.  
\end{equation} 
This process corresponds to the absorption of both photons by an atom 
before emission of either, and thus describes a one-atom process which
is of the same order in the atomic number density as one-atom 
scattering.
The kinematical and geometric factors of Eq. (\ref{OneAtomEff}) 
Eq. (\ref{Geometrics}) are the same for this process, and the 
resonance denominators are
\begin{eqnarray}
\label{ResDenomsLadder}
R^{(3)}_{1} & = &
c(k+l-k') -\frac{\hbar}{M}[{\bf p} \cdot {\bf l}' + l'^{2}/2]
- \omega_{da} + i \gamma_{d}
\nonumber \\
R^{(3)}_{2} & = &
c(k+l) - \frac{\hbar}{M}[{\bf p} \cdot ({\bf k}+{\bf l}) + |{\bf k} + {\bf l}|^{2}/2]
\nonumber \\
& & - \omega_{ca} +  i \gamma_{c} 
\nonumber \\
R^{(3)}_{3} & = &
c(k) -\frac{\hbar}{M}[{\bf p} \cdot {\bf k} + k^{2}/2]
- \omega_{ba} + i \gamma_{b}.
\end{eqnarray}


Right-angle scattering amplitudes for this process are shown in Fig.  
10.
Parameters are as for Fig. 2.

\subsection{Lorentz-model Behavior}

It is interesting to consider the case of a ladder atom with equal 
energy spacing $\omega_{cb}=\omega_{ba}$ and matrix elements 
$|\mu_{cb}|^{2}=2|\mu_{ba}|^{2}$.  In this case the states $a$--$c$ 
are equivalent to the lowest three levels of a harmonic oscillator, 
i.e., to a Lorentz model, and the medium is effectively linear for 
two-photon processes.


The amplitudes for the one atom process of Eq. (\ref{OneAtomProcess})
and the 
ladder process of Eq. (\ref{LadderProcess}) partially cancel.  
The resulting 
signal is smaller and lacks oscillations, as shown in Fig.  11.
Parameters are as for Fig. 2.

\subsection{Background Events}

In addition to the photon-photon scattering processes, Rayleigh 
scattering (and Raman scattering for more complicated atoms) will 
create an uncorrelated coincidence background.  This background is 
calculated in Appendix \ref{Correlations}.  The coincidence signal, 
consisting of both the Lorentz-model atom photon-photon scattering signal and 
the incoherent background is shown in Fig.  12.
 The peak 
coincidence rate (at $\delta \tau = 0$) is approximately twice the 
background, accidental coincidence rate.  In the limit of large 
detuning, it becomes exactly twice accidental rate.  This can be 
explained in analogy with the Hanbury-Brown-Twiss effect as follows: 
For the optimal geometry the drive beams are conjugates of each other 
$H({\bf x}) = G^{*}({\bf x})$ and the detectors are in opposite 
directions.  The linear atoms act to create a random index grating 
which scatters a chaotic but equal (up to phase conjugation) field to 
each detector.  As expected for chaotic light \cite{Walls=1994}, the 
fourth-order equal-time correlation function is twice the product of 
second-order correlation functions.
\begin{equation}
\left<E^{2}({\bf x}_{A},t)E^{2}({\bf x}_{B},t)\right> = 2 
\left<E^{2}({\bf x}_{A},t)\right>\left<E^{2}({\bf x}_{B},t)\right>.  
\end{equation}

\section{Small-angle Scattering}

%

Thus far the discussion has involved only large-angle scattering.  In 
the context of cavity nonlinear optics all fields are propagating 
nearly along the optical axis of the cavity so it is necessary
to consider scattering processes for nearly co-propagating or nearly 
counter-propagating photons.  As argued above, the temporal width of 
the correlation signal scales as $1/q$, the inverse of the momentum 
transfer.  This is shown in Fig. 13
and 
Fig. 14,
which show rates for scattering photons from beams in 
the $x$--$z$ plane into the the $y$--$z$ plane.  In all cases the beam 
directions are $0.1$ radian from the $z$ axis.  The coincidence 
distribution shows oscillations which die out on the time-scale of the 
inverse Doppler width, and a non-oscillating pedestal with a width 
determined by the momentum transfer $q$.

%

The pedestal, however, does not correspond to the duration of the 
nonlinear process in this case.  As above, by considering a ladder 
atom with the energy spacings and matrix elements of a harmonic 
oscillator we can isolate the linear optical behavior.  As shown in 
Fig. 15
and Fig. 16,
this behavior includes 
the pedestal, but not the oscillations, indicating that the nonlinear 
optical process is still fast, with a time-scale on the order of the 
inverse Doppler width.

\section{Limitations on scattering angle}

Due to the limited width of the atomic momentum distribution, the 
resonance denominator $R^{(1)}_{2}$ is small if the input and output 
photons are not of nearly the same energy.  Since the complete process 
must conserve photon momentum, input photons with net transverse 
momentum in the output photon direction will scatter less strongly.  
The width of this resonance is very narrow: a net transverse momentum 
$k_{y} + l_{y} \sim k \sqrt{k_{\rm B}T/Mc^{2}}$ is sufficient that few 
atoms will be resonant.  As $\sqrt{k_{\rm B}T/Mc^{2}}$ is typically of 
order $10^{-6}$ in an atomic vapor, this would be a severe restriction 
on the transverse momentum content of the beams in a cavity nonlinear 
optics experiment.  However, as shown in Fig.  16,
the 
narrow resonance associated with $R^{(1)}_{2}$ contributes the linear 
response of the medium.  The nonlinear response, which has the same 
resonance
character as the ``ladder'' process, is not limited in this way 
because $R^{(3)}_{2}$ does not depend upon the output photon energies.

\section{Output Polarization}

The polarization of the output photons depends on the structure of the 
atom and can produce polarization-entangled photons.  For example, if 
the input photons are propagating in the $\pm z$ directions and are $x$ 
polarized, the two absorption events in the above diagram change the 
$z$ component of angular momentum by $\delta m = \pm 1$.  In order for 
the process to return the atom to its initial state, the two emission 
events must both produce $\delta m = \pm 1$ or both $\delta m = 0$.  
For right angle scattering with the detectors in the $\pm y$ 
directions, the output photons must therefore be either both $x$ or 
both $z$ polarized.  If both polarizations are possible, the emitted 
photons are entangled in polarization, as well as in energy and in 
momentum.

\section{Conclusion}

Time correlations in photon-photon scattering provide an indication of 
the time-scale over which the atomic medium is involved in the 
interaction among photons in a nonlinear medium.  It is found that the 
time-scale is determined by the inhomogeneous broadening of the medium 
and the magnitude of the momentum transfer.  For large-angle 
scattering, the time-scale of involvement is $\delta \tau \sim \Lambda 
M/\hbar q$, while for small-angle scattering the time-scale is $\delta 
\tau \sim \Lambda M/\hbar k$.  As this time-scale is shorter than the 
atomic relaxation time, calculations which contain an adiabatic 
elimination of the atomic degrees of freedom necessarily overlook the 
fastest dynamics in this process.  

\appendix

\section{Photon Correlations}
\label{Correlations}
\subsection{Detection Amplitudes}

Unlike a genuine two-body collision process, atom-mediated 
photon-photon scattering has a preferred reference frame which is 
determined by the atomic momentum distribution.  To calculate 
the photon correlations we work in the 
``laboratory'' frame and assume the momentum distribution is 
symmetric about zero.  We consider scattering from two input beams 
with beam shapes
$G({\bf x}) \equiv {V^{-1/2}}\sum_{{\bf k}} g({\bf k}) 
\exp[{i{\bf k}\cdot{\bf x}} ]$ and $
H({\bf x}) \equiv {V^{-1/2}}\sum_{{\bf l}} h({\bf l}) 
\exp[{i{\bf l}\cdot{\bf x}} ]$ which are normalized as
$
\sum_{{\bf k}} | g({\bf k})|^{2}= \sum_{{\bf l}} |h({\bf l})|^{2} = 1$.
We further assume that the beams are derived from the same 
monochromatic source and are paraxial, i.e., that $g({\bf k})$ 
is only appreciable in some small neighborhood of the average beam 
direction ${\bf k}_{0}$, and similarly for $h({\bf l})$ around 
${\bf l}_{0}$.   The geometry is shown schematically in Fig. 17.
For convenience, the beams are assumed to each contain one photon, so 
that the initial state of the field is
\begin{equation}
\left|\phi(0)\right> = A_{G}^{\dag}A_{H}^{\dag}\left|0\right>
\end{equation}
where the creation operators $A_{G}^{\dag},A_{H}^{\dag}$ are
$A_{G}^{\dag} \equiv \sum_{{\bf k}} g({\bf k}) a_{{\bf k}}^{\dag}$
and
$A_{H}^{\dag} \equiv \sum_{{\bf l}} h({\bf l}) a_{{\bf l}}^{\dag}.$
Scaling of the result to multiple photons is obvious.



We use Glauber photodetection theory to determine the rates at which 
scattering products arrive at two detectors $A$ and $B$ at space-time 
points $({\bf x}_{A},t_{A})$ and $({\bf x}_{B},t_{B})$, respectively.  
We compute the correlation function in the Heisenberg representation
\begin{eqnarray}
\label{CorrFun}
\makebox[0.5cm][l]{$\displaystyle
P({\bf x}_{A},t_{A},{\bf x}_{B},t_{B})
$} & & 
\nonumber \\
& = & 
|\left<0\right|\Phi^{(+)}_{\rm H}({\bf x}_{B},t_{B})
\Phi^{(+)}_{\rm H}({\bf x}_{A},t_{A})
\left|\phi(0)\right>_{\rm H}|^{2}
\end{eqnarray}
where the photon field operator is 
\begin{equation}
\Phi^{(+)}_{\rm H}({\bf x},t) \equiv 
V^{-1/2}\sum_{{\bf k},\alpha}a_{{\bf k},\alpha}(t)
\exp[i{\bf k}\cdot{\bf x}].
\end{equation}
This field operator is similar to the positive frequency part of the 
electric field and is chosen so that 
$\Phi^{(-)}({\bf x},t)\Phi^{(+)}({\bf x},t)$ is Mandel's photon-density 
operator \cite{Mandel=1995}.  To make use of perturbation theory, 
Eq. (\ref{CorrFun}) is more conveniently expressed in interaction 
representation as
\newcommand{\ampl}{{A}}
\begin{eqnarray}
\makebox[0.5cm][l]{$\displaystyle
P({\bf x}_{A},t_{A},{\bf x}_{B},t_{B})
$} & & 
\nonumber \\
& = & 
|\left<0\right|\Phi^{(+)}_{\rm I}({\bf x}_{B},t_{B})
U_{I}(t_{B},t_{A})
\Phi^{(+)}_{\rm I}({\bf x}_{A},t_{A})
\left|\phi(t_{A})\right>_{\rm I}|^{2}
\nonumber \\
& = & 
|\left<0\right|\Phi^{(+)}_{\rm I}({\bf x}_{B},t_{B})
\Phi^{(+)}_{\rm I}({\bf x}_{A},t_{A})
\left|\phi(t_{A})\right>_{\rm I}|^{2}
\nonumber \\
& \equiv & 
|\ampl({\bf x}_{A},t_{A},{\bf x}_{B},t_{B})|^{2}
\end{eqnarray}
where $U_{I}$ is the interaction picture time-evolution operator,
the interaction picture field operator is 
\begin{equation}
\Phi^{(+)}_{\rm I}({\bf x},t)
= V^{-1/2}\sum_{{\bf k},\alpha}a_{{\bf k},\alpha} 
\exp[i({\bf k}\cdot{\bf x}-ckt)]
\end{equation} 
and in passing to the second line we have made the assumption that a 
detection at $({\bf x}_{A},t_{A})$ does not physically influence the 
behavior of photons at $({\bf x}_{B},t_{B})$ although there may be 
correlations.  
The the amplitude of joint detection is 
\begin{eqnarray}
\makebox[0.5cm][l]{$\displaystyle
\ampl({\bf x}_{A},t_{A},{\bf x}_{B},t_{B})
$} & & 
\nonumber \\
& = & \frac{(2\pi)^{3}}{V^{2}\hbar} \sum_{{\bf k}'{\bf l}'} 
\exp[i({\bf k}'\cdot{\bf x}_{A}-c k't_{A})]
\nonumber \\
& & \times 
\exp[i({\bf l}'\cdot{\bf x}_{B}-c l't_{B})]
\nonumber \\
& & \times 
\sum_{{\bf k}{\bf l}}g({\bf k})h({\bf l}) 
V_{{\bf l}'{\bf k}'{\bf l}{\bf k}}
\nonumber \\
& & \times 
\frac{1-\exp[ic(k'+l'-k-l)t_{A}]}
     {c(k'+l'-k-l)+i \eta}
\end{eqnarray}

Although $V_{{\bf l}'{\bf k}'{\bf l}{\bf k}}$ depends strongly upon the 
magnitudes of the initial and final photon momenta through the 
resonance denominators of Eq. (\ref{ResDenomsOne}), it depends only 
weakly on their directions through the geometrical factors of 
Eq. (\ref{Geometrics}).  This and the assumption of 
paraxial input beams justify the approximation
\begin{eqnarray}
\makebox[0.5cm][l]{$\displaystyle
\sum_{{\bf k}{\bf l}} g({\bf k})h({\bf l}) V_{{\bf l}'{\bf k}'{\bf l}{\bf k}} 
$} & & 
\nonumber \\
&\approx &
V_{{\bf l}'{\bf k}'{\bf l}_{0}{\bf k}_{0}} 
\sum_{{\bf k}{\bf l}} g({\bf k})h({\bf l}) 
\delta_{{\bf k}+{\bf l},{\bf k}'+{\bf l}'}
\nonumber \\
& = & V_{{\bf l}'{\bf k}'{\bf l}_{0}{\bf k}_{0}}
\int d^{3}{\bf x} G({\bf x})H({\bf x}) \exp[-i({\bf k}'+{\bf l}')\cdot {\bf x}] .
\end{eqnarray}


We can similarly treat the output photons in the paraxial 
approximation for the case that the detection points are far from the 
interaction region, i.e., that $x_{A},x_{B} \gg x$.
Making these approximations and dropping unphysical portions of the 
solution propagating inward from the detectors toward the source region, 
we find
%
%
\begin{eqnarray}
\makebox[0.5cm][l]{$\displaystyle
\ampl({\bf x}_{A},t_{A},{\bf x}_{B},t_{B})
$} & & 
\nonumber \\
& = & \frac{-i}{\hbar c} \int k'd k'l'
V_{{\bf l}'{\bf k}'{\bf l}_{0}{\bf k}_{0}}
\nonumber \\
& & \times 
\int d^{3}{\bf x} \frac{G({\bf x})H({\bf x})}{|{\bf x}_{A}-{\bf x}||{\bf x}_{B}-{\bf x}|}
\nonumber \\
& & \times 
\exp[i({\bf k}'\cdot({\bf x}_{A}-{\bf x})-c k't_{A})]
\nonumber \\
& & \times 
\exp[i({\bf l}'\cdot({\bf x}_{B}-{\bf x})-c l't_{B})]
\nonumber \\
& & \times 
\theta(\tau_{A}) \theta(\tau_{B})
\end{eqnarray}
where $c\tau_{A,B}\equiv ct_{A,B}-x_{A,B}$ are retarded times.

A final approximation ignores the slow variation of $k',l'$ 
relative to that of the resonant $V_{{\bf l}'{\bf k}'{\bf l}_{0}{\bf k}_{0}}$.  
Further, we define $G'({\bf x}) \equiv G({\bf x}) \exp[ik_{0}\cdot {\bf x}]$,
$H'({\bf x}) \equiv H({\bf x}) \exp[il_{0}\cdot {\bf x}]$ and 
$k' \equiv k'_{0} + \delta k'$ where ${\bf k}'_{0}$ is 
the value of ${\bf k}'$ which maximizes $V_{{\bf l}'{\bf k}'{\bf l}_{0}{\bf k}_{0}}$ 
subject to momentum and energy conservation.  
This gives a simple expression for the correlation function 
\begin{eqnarray}
\makebox[0.5cm][l]{$\displaystyle
\ampl({\bf x}_{A},t_{A},{\bf x}_{B},t_{B})
$} & & 
\nonumber \\
&= & \frac{-ik'l'}{\hbar c} 
\exp[-ic(k'_{0}\tau_{A}+l'_{0}\tau_{B})]
\nonumber \\
& & \times
\int d \delta_{k'} V_{{\bf l}'{\bf k}'{\bf l}_{0}{\bf k}_{0}} 
\exp[i c \delta_{k'}(\tau_{B}-\tau_{A})] 
\nonumber \\
& & \times
\int d^{3}{\bf x} \frac{G'({\bf x})H'({\bf x})}{|{\bf x}_{A}-{\bf x}||{\bf x}_{B}-{\bf x}|}
\nonumber \\
& & \times
\exp[i({\bf k}_{0}+{\bf l}_{0}-{\bf k}'_{0} - {\bf l}'_{0})\cdot {\bf x}]
\nonumber \\
& & \times 
\theta(\tau_{A}) \theta(\tau_{B})
.
\end{eqnarray}
This can be interpreted as consisting of a carrier wave, a Fourier 
transform of the scattering amplitude and a coherent integration of 
the contributions from different parts of the interaction region.  
The spatial integral enforces phase matching in the photon-photon 
scattering process.

\subsection{Detection Rates}
\index{detection rate}

The probability for a coincidence detection at two detectors of 
specified area and 
in two specified time intervals is
\begin{equation}
P = \int d^{2}{\bf x}_{A} d^{2}{\bf x}_{B} c dt_{A} c dt_{B}
|\ampl({\bf x}_{A},t_{A},{\bf x}_{B},t_{B})|^{2},
\end{equation}
where the integral is over the detector surfaces (each assumed normal to the 
line from scattering region to detector) and over the relevant time 
intervals.
This is more conveniently expressed in terms of a rate $W$ of 
coincidence detections in terms of the detector solid angles 
$\delta\Omega_{A}$, $\delta\Omega_{B}$ and the 
difference in retarded arrival times 
$\tau_{-} \equiv \tau_{B}-\tau_{A}$
\begin{equation}
W = c^{2}x_{A}^{2}x_{B}^{2}
|\ampl({\bf x}_{A},t_{A},{\bf x}_{B},t_{B})|^{2}
\delta\Omega_{A}\delta\Omega_{B}d\tau_{-}.
\end{equation}

Coincidence rate is largest when the detectors 
are placed in the directions which satisfy the phase-matching condition.
We assume that ${\bf k} + {\bf l} = {\bf k}' + {\bf l}' = 0$ and
that the detectors are small compared to the source-detector 
distance, i.e., that $\delta\Omega_{A,B}\ll 1$.  Under these 
conditions, the rate of coincidence events reduces to 
\begin{eqnarray}
\label{RateEqn}
W_{\rm scattering} &=& \frac{(k'l')^{2}}{\hbar^{2}}
\left|\int d \delta_{k'} V_{{\bf l}'{\bf k}'{\bf l}_{0}{\bf k}_{0}} 
\exp[i c \delta_{k'}\tau_{-}]\right|^{2}
\nonumber \\
& & \times 
\left|\int d^{3}x 
G({\bf x})H({\bf x})\right|^{2}
\delta\Omega_{B}\delta\Omega_{B}d\tau_{-}.
\end{eqnarray} 

\subsection{Signal Contrast}

\index{signal contrast}
\index{scattering!Rayleigh and Raman}
In addition to the photon-photon scattering signal, uncorrelated 
Rayleigh and Raman scattering events give a background of accidental 
coincidences.  
The rate of scattering into a 
small solid angle $\delta\Omega$ is
\begin{equation}
W_{\rm BG} = B \delta\Omega \int d^{3}x n_{k} 
\end{equation}
where
\begin{eqnarray}
B &\equiv &
  \sum_{a,c}\int d^{3}{\bf p}\left<n_{a,{\bf p}}\right>
  \left(1\pm\left<n_{c,{\bf p}'}\right>\right) 
  \frac{k_{f}^{4}c}{(2\pi)^{3}\hbar^{2}} 
\nonumber \\
& & \times
\left| \sum_{b}
\frac{ ({\bf e}_{f}\cdot{\mbox{\boldmath $\mu$}}_{bc})^{*}{\bf e}_{i}\cdot{\mbox{\boldmath $\mu$}}_{ba}}
{ck + \omega_{ab}-\frac{\hbar}{M}[{\bf p}\cdot{\bf k} + k^{2}/2] + i 
\gamma_{b}} \right|^{2} 
\end{eqnarray}
and $n_{k}$ is the number density of photons propagating in the ${\bf 
k}$ direction.  
In terms of the beam-shape functions for two colliding beams, the rate
of accidental coincidences is 
\begin{eqnarray}
W_{\rm accidental} &=& B^{2} \left[ \int d^{3} x |G({\bf x})|^{2} + 
|H({\bf x})|^{2}\right]^{2} 
\nonumber \\ & & \times
\delta\Omega_{A}\delta\Omega_{B}d\tau_{-}.
\end{eqnarray}
The ratio of coincidences 
due to photon-photon 
scattering to accidental background coincidences is thus
\begin{eqnarray}
\label{IdealContrast}
\frac{W_{\rm scattering}}{W_{\rm accidental}} &= &
\frac{(k'l')^{2}}{4\hbar^{2}}
\frac{F}{B^{2}}
\nonumber \\
& & \times
\left|\int d \delta_{k'} V_{{\bf l}'{\bf k}'{\bf l}_{0}{\bf k}_{0}} 
\exp[i c \delta_{k'}\tau_{-}]\right|^{2} 
\end{eqnarray}
where $F$ is the mode fidelity factor 
\begin{equation} 
\label{ModeFactor}
F  \equiv  4
\frac{
\left[\int d^{3}x G({\bf x})H({\bf x})\right]^{2}
}{
\left[\int d^{3} x \left(|G({\bf x})|^{2} + |H({\bf x})|^{2}\right)\right]^{2}
}.
\end{equation}

\bibliography{PPSA}
\bibliographystyle{prsty}

\end{multicols}

\begin{figure}
\centerline{\psfig{width=2in,figure=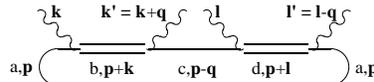}}
\label{OneAtomFig} 
\caption{FIG. 1. Photon-photon scattering at a single atom.}
\end{figure}

\begin{figure}
\begin{center}
\epsfig{file=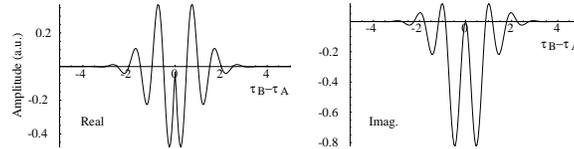,width=3.0in,angle=0}
\caption{FIG. 2.  Right-angle scattering amplitude $A$ vs. time delay for the 
single-atom process of Fig. 1.
The time unit is $\delta_{\rm D}^{-1}$.}
\label{OARAReImFig}
\end{center}
\end{figure}

\begin{figure}
\begin{center}
\epsfig{file=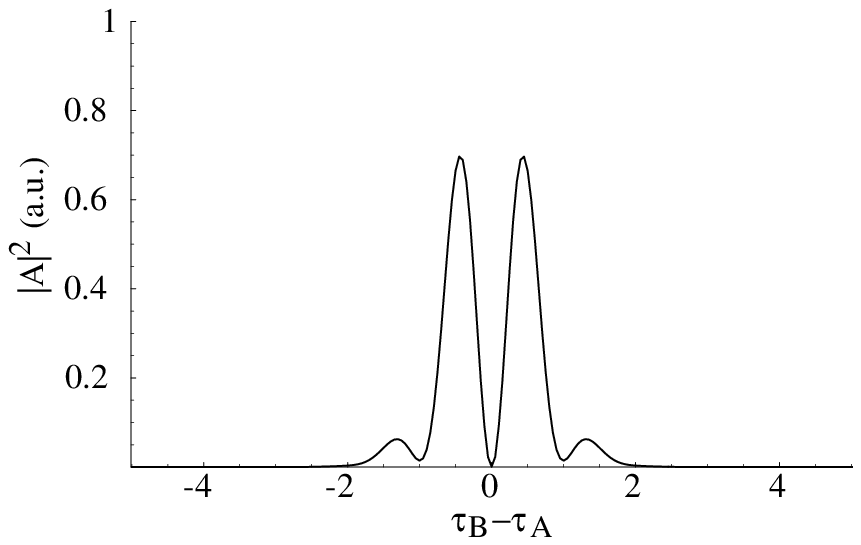,width=2.5in,angle=0} 
\caption{FIG. 3.  Right-angle 
scattering rate $|A|^{2}$ vs.  time delay for the single-atom process 
of Fig.  1.
Time unit is $\delta_{\rm D}^{-1}$.}
\label{OARAPwFig}
\end{center}
\end{figure}

\begin{figure}
\begin{center}
\epsfig{file=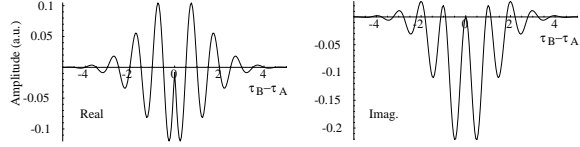,width=3.0in,angle=0}
\caption{FIG. 4.  Statistical correction to the one-atom scattering 
amplitude.  The time unit is $\delta_{\rm D}^{-1}$.}
\label{SCRAReImFig}
\end{center}
\end{figure}

\begin{figure}
\centerline{\psfig{width=1.5in,figure=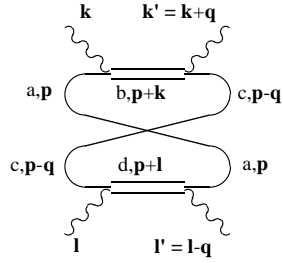}}
\label{TwoAtomFig} 
\caption{FIG. 5.  Two-atom photon-photon scattering.}
\end{figure}

\begin{figure}
\begin{center}
\epsfig{file=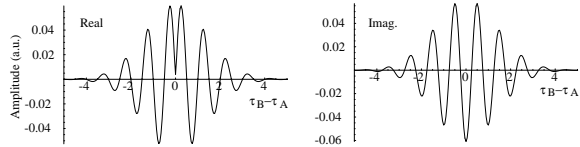,width=3.0in,angle=0}
\caption{FIG. 6.  Scattering rate $|A|^{2}$ vs. time delay for the two-atom 
process of Fig. 5.
The time unit is $\delta_{\rm D}^{-1}$.}
\label{TARAReImFig}
\end{center}
\end{figure}

\begin{figure}
\begin{center}
\epsfig{file=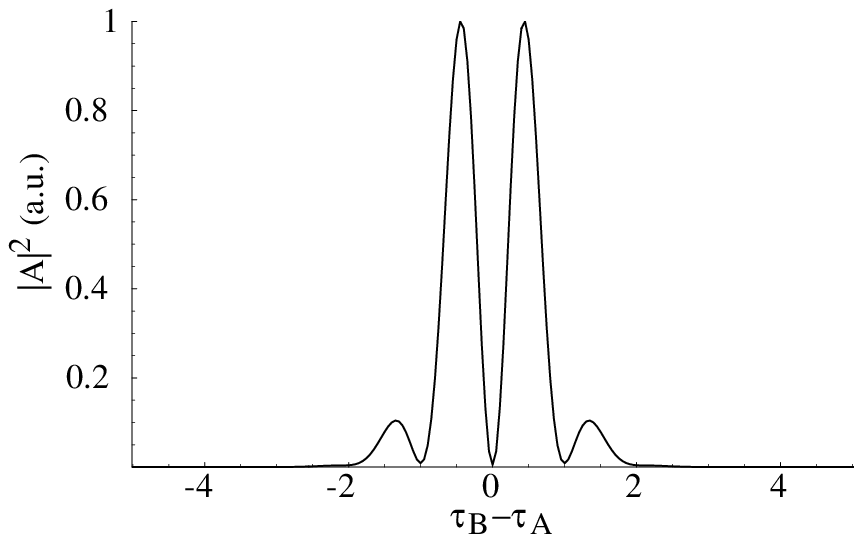,width=2.5in,angle=0}
\caption{FIG. 7.  Scattering rate $|A|^{2}$ vs. time delay for a Bose gas of 
phase-space density 1/2.  The time unit is $\delta_{\rm D}^{-1}$.}
\label{BORAPwFig}
\end{center}
\end{figure}

\begin{figure}
\begin{center}
\epsfig{file=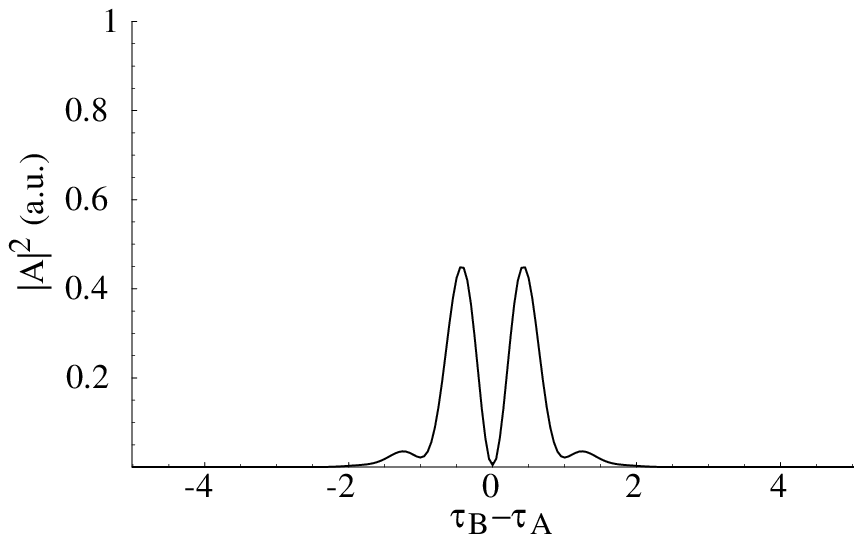,width=2.5in,angle=0}
\caption{FIG. 8.  Scattering rate $|A|^{2}$ vs. time delay for a Fermi gas of 
phase-space density 1/2.  The time unit 
 is $\delta_{\rm D}^{-1}$.}
\label{FERAPwFig}
\end{center}
\end{figure}

\begin{figure}
\centerline{\psfig{width=2in,figure=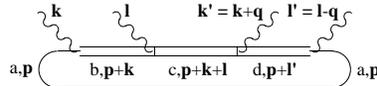}}
\label{LadderAtomFig} 
\caption{FIG. 9.  ``Ladder'' process in a three-level atom.}
\end{figure}

\begin{figure}
\begin{center}
\epsfig{file=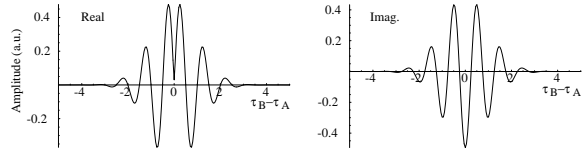,width=3.0in,angle=0}
\caption{FIG. 10.  Scattering rate $|A|^{2}$ vs. time delay for the ``ladder'' 
process of Fig. 9.
The time unit is $\delta_{\rm D}^{-1}$.}
\label{LARAReImFig}
\end{center}
\end{figure}

\begin{figure}
\begin{center}
\epsfig{file=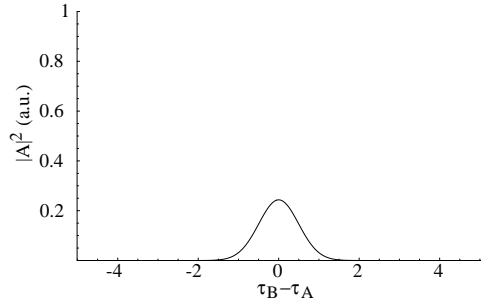,width=2.5in,angle=0}
\caption{FIG. 11.  Scattering rate $|A|^{2}$ vs. time delay for a 
Lorentz-model atomic medium.  The time unit is $\delta_{\rm D}^{-1}$.}
\label{LIRAPwFig}
\end{center}
\end{figure}

\begin{figure}
\begin{center}
\epsfig{file=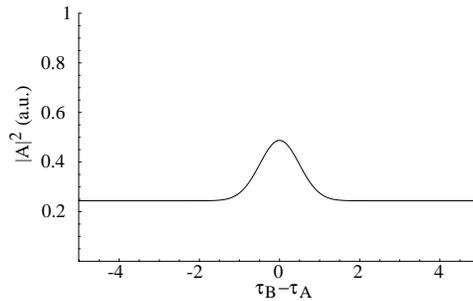,width=2.5in,angle=0}
\caption{FIG. 12.  Coincidence rate vs. time delay for a Lorentz-model 
atomic medium.  The constant background is accidental coincidences due 
to independent Rayleigh scattering events.
The time unit is $\delta_{\rm D}^{-1}$.}
\label{LIRABgFig}
\end{center}
\end{figure}

\begin{figure}
\begin{center}
\epsfig{file=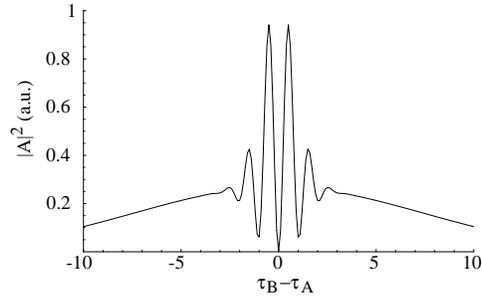,width=2.5in,angle=0}
\caption{FIG. 13.  Small angle scattering rate $|A|^{2}$ vs. time delay for 
nearly co-propagating photons.
Time unit is $\delta_{\rm D}^{-1}$.  }
\label{OACOPwFig}
\end{center}
\end{figure}

\begin{figure}
\begin{center}
\epsfig{file=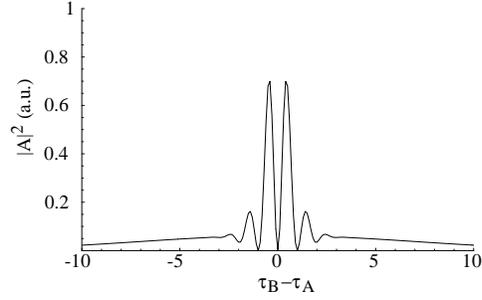,width=2.5in,angle=0}
\caption{FIG. 14.  Right-angle scattering rate $|A|^{2}$ vs. time delay for 
nearly counter-propagating photons.
Time unit is $\delta_{\rm D}^{-1}$.}
\label{OACTPwFig}
\end{center}
\end{figure}

\begin{figure}
\begin{center}
\epsfig{file=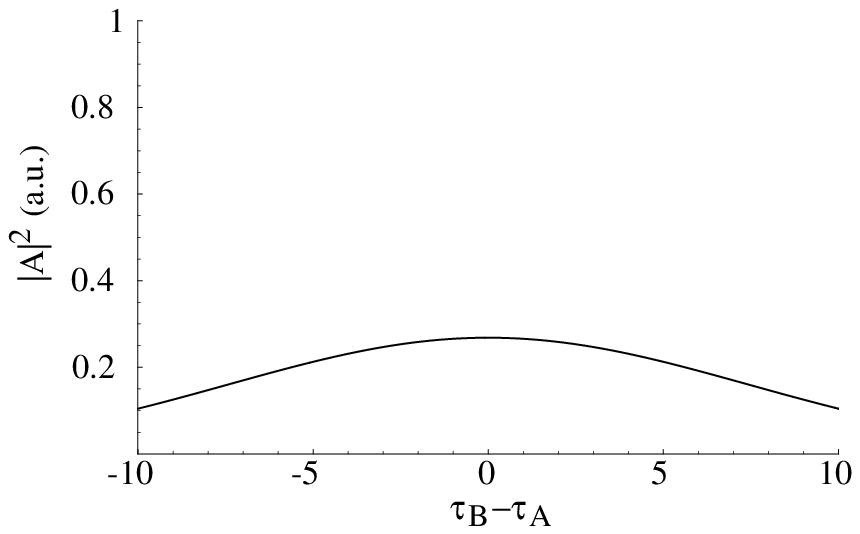,width=2.5in,angle=0}
\caption{FIG. 15.  Coincidence rate $|A|^{2}$ vs. time delay for 
nearly co-propagating photons in a linear medium.
$\gamma_{b} = 0.01 \delta_{\rm D}$, $\Delta = 2 \pi \delta_{\rm D}$.  
Time unit is $\delta_{\rm D}^{-1}$.  }
\label{LICOPwFig}
\end{center}
\end{figure}

\begin{figure}
\begin{center}
\epsfig{file=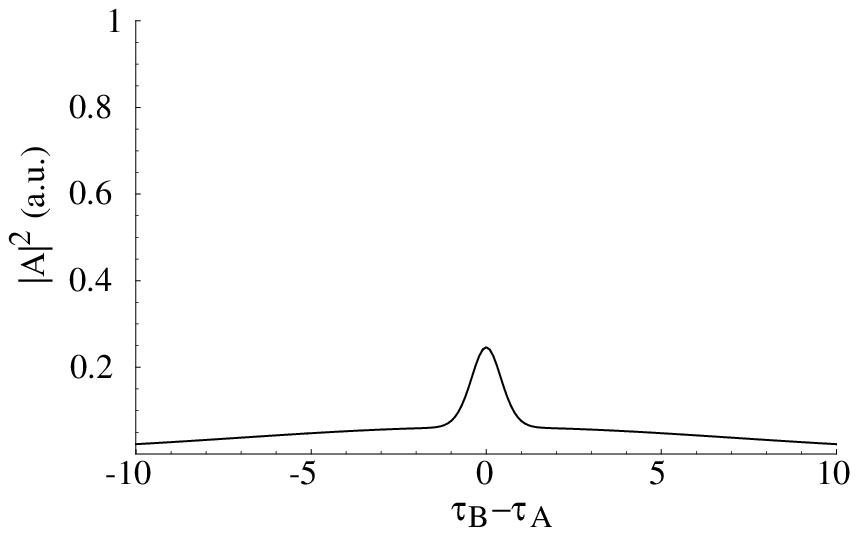,width=2.5in,angle=0}
\caption{FIG. 16.  Coincidence rate $|A|^{2}$ vs. time delay for 
nearly counter-propagating photons linear medium.
$\gamma_{b} = 0.01 \delta_{\rm D}$, $\Delta = 2 \pi \delta_{\rm D}$.  
Time unit is $\delta_{\rm D}^{-1}$.}
\label{LICTPwFig}
\end{center}
\end{figure}

\begin{figure}
\centerline{\psfig{width=2.5in,figure=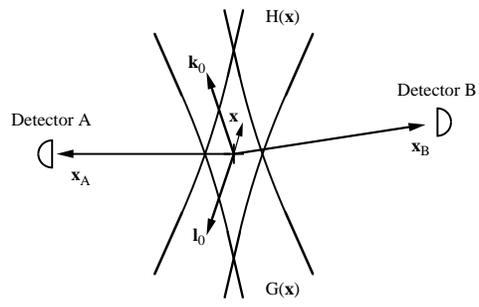}}
\label{BeamGeomFig} 
\caption{FIG.  17.  Geometry of collision process.}
\end{figure}

\end{document}